# Study of polymer-solvent interactions of complex polysiloxanes using dissipative particle dynamics

Javier Vallejo-Montesinos[a1], Antonio Villegas[a2], Jorge Cervantes[a3], Elías Pérez[b] and Armando Gama Goicochea[c†]

[a]División de Ciencias Naturales y Exactas, Universidad de Guanajuato, Campus Guanajuato, Col. Noria Alta s/n, Guanajuato 36050, Guanajuato, Mexico

[b]Instituto de Física, Universidad Autónoma de San Luis Potosí, Álvaro Obregón 64, , San Luis Potosí 78000, Mexico

[c]División de Ingeniería Química y Bioquímica, Tecnológico de Estudios Superiores de Ecatepec, Av. Tecnológico s/n, Ecatepec 55210, Estado de México, Mexico

## ABSTRACT

The mesoscopic modeling of three polysiloxanes in solution is reported in this work, with the purpose of predicting their physicochemical properties as functions of the quality of the solvent, so that a judicious choice of polymer/solvent can be made for various applications. The polymers studied were the following polysiloxanes: polydimethylsiloxane (PDMS), polysiloxane with a bulky alkyl side group (PMHS) and a siloxane copolymer with a hydrophilic polar side group (P2DMPAS). The model used and solved through numerical simulations is the one known as dissipative particle dynamics. Density profiles and radial distribution functions were calculated for each system. We analyzed how the polymers behave in the presence of solvents of varying quality and compared their behavior with experimental data. We observed that we could replicate the behavior in good solvents for PDMS and PMHS. We also observed in the simulation box the formation of pseudo-micelles for P2DMPAS.

Electronic mails: [a1]javas210@ugto.mx; [a2]vigaja@ugto.mx; [a3]jauregi@ugto.mx; [b]elias@ifisica.uaslp.mx; [c†]agama@alumni.stanford.edu (Corresponding author).

**Introduction**

Polysiloxanes are materials with a wide range of applications, largely due to their desirable properties, such as high flexibility, thermal resistance, and exposure to UV light resistance, among others [1,2]. Since they are often polymerized in and processed from solution, it is very important to understand their interactions and associations with diverse solvents. For these materials, multiple synthetic routes can be used [1,2]; however, aside from polydimethylsiloxane, little has been reported regarding solution behavior for this family of polymers. Changing the side group in a polysiloxane can change its solubility in various solvents, or lead even to unusual behavior, as we have reported previously [3–6]. For example, we reported earlier [4] the analysis of the second virial coefficient, $A_2$, for a series of polysiloxane chains with different side groups; the effect of the side group and molecular weight on the value of $A_2$ was analyzed through the Helical Worm (HW) chain type model [7]. The theoretical analysis of the interpenetration function "$\psi$" is an excellent way to measure excluded volume effects and, for this case, it was performed using the Two Parameters (TP) scheme [7]. The theoretical and experimental behavior of $A_2$ for a series of polysiloxanes was investigated [4] by considering the effects of the type of side group attached to the main chain on the molecular conformation and the molecular weight of the polymer. Our results in that study showed a strong influence of the flexibility of the side groups and the molecular weight of the polymer on the structural properties of the fluid. We reported that polysiloxanes with very large side groups exhibited a significant difference in the theoretical value of $A_2$ obtained through the TP scheme relative to their experimental value.

Given the unusual behavior presented by these asymmetrical (the asymmetry, in that case, was with respect to the difference in size of the two types of side groups attached to the silicon atoms) short chains of polysiloxanes with a bulky side group (hexyl and hexadecyl), we made a subsequent theoretical study in which we analyzed the behavior of the $A_2$ and the experimental values of the molecular conformations (obtained from the slope of the graph of the radius of gyration against molecular weight) in a good solvent for short chains of poly-n-hexylmethylsiloxane (PMHS) [5]. We also considered a series of copolysiloxanes with a hydrophilic lateral group, exhibiting good solubility in ethanol with no solubility at all in water and poor solubility in toluene for a copolymer which was mostly polydimethylsiloxane (PDMS) [6]. Even PDMS, by itself, exhibits remarkable features when it is studied in solution, like the draining effect [4, 8]. These unusual properties in solution are closely related to the inner properties of the siloxane bond (such as a bond angle of 143°, near zero value of torsional barrier, etc.), which is a very flexible bond [1, 9]. To better understand phenomena such as this, simulation techniques are becoming very important research tools. In this work, we used those tools to specifically study the influence of the quality of the solvent on the structural and thermodynamic properties of several complex polysiloxanes of current interest. Most previously reported works have focused on the properties of these polymers and their mutual association, but so far, there are few reports that systematically study the influence of the solvent's characteristics on the polymer properties, such as $A_2$, radius of gyration ($RMS_{radius}$), radial distribution function, etc. That is the central focus of this work and, as we shall show here, modeling the solvent quality can give rise to complex polymer associations.

More than twenty years ago Hoogerbrugge and Koelman [10, 11] introduced a new simulation technique to study the hydrodynamic behavior of colloidal dispersions, which they called dissipative particle dynamics (DPD). This technique is based on the simulation of soft spheres, whose motion is governed by simple, linearly decaying, conservative forces. By introducing bead-and-spring type particles, polymers can be simulated with the same method. Español and Warren [12] later developed the statistical mechanics of DPD, which incorporated the fluctuation-dissipation theorem into the DPD model. In the DPD formulation, all particles interact through three forces: a conservative force $F^C$, a dissipative force $F^D$, and a random force $F^R$ [10–28]. This kind of simulation opens up the way to bridge the gap from atomistic simulations, where solubility parameters can be calculated, to mesoscopic simulations where mesophases and network formation can be studied. DPD simulations have been employed successfully for the prediction of the properties of various polymeric systems [13–15, 17–21, 27, 29, 30]. In DPD, the fluid is coarse-grained into soft, momentum carrying beads, with their dynamics governed by the following stochastic differential equations represented in Eqs. (1-3):

$$\frac{dv_i}{dt} = F_{ij}^C + F_{ij}^D + F_{ij}^R \tag{1}$$

$$\frac{dr_i}{dt} = v_i \tag{2}$$

The conservative force $F^C$, dissipative force $F^D$, and random force $F^R$ are defined as:

$$\begin{aligned} F_{ij}^C &= a_{ij}\omega^C(r_{ij})\hat{e}_{ij} \\ F_{ij}^D &= -\gamma\omega^D(r_{ij})[\hat{e}_{ij}\cdot\vec{v}_{ij}]\hat{e}_{ij} \\ F_{ij}^R &= \sigma\omega^R(r_{ij})\hat{e}_{ij}\xi_{ij} \end{aligned} \tag{3}$$

where $r_{ij}$ is the relative position vector, defined as $r_{ij}= r_i-r_j$, $\hat{e}_{ij}$ is the unit vector in the position $r_{ij}$, and it is defined as $\hat{e}_{ij} = r_{ij}/r_{ij}$ where $r_{ij}$ is the distance between particles $i$ and $j$, $r_i$ is the position of particle $i$, and $v_{ij} = v_i-v_j$ with $v_i$ being the velocity of particle $i$, as in Eq. (1). The variable $\xi_{ij}$ is a random number uniformly distributed between 0 and 1 with a Gaussian distribution and unit variance; $a_{ij}$, $\gamma$, and $\sigma$ are the strengths of the conservative, dissipative, and random forces, respectively. $\omega^C(r_{ij})$ and $\omega^R(r_{ij})$ are dimensionless weight functions given by

$$\omega^C(r_{ij}) = \omega^R(r_{ij}) \equiv \omega(r_{ij}) \tag{4}$$

and

$$\omega(r_{ij}) = \begin{cases} 1 - r_{ij}/R_C, & r_{ij} \leq R_C \\ 0, & r_{ij} \geq R_C \end{cases} \tag{5}$$

where $R_c$ is a cutoff distance. It should be remarked that this choice of distance – dependent conservative force is not without foundation, as it has been shown to arise from properly averaged, microscopic interactions, such as the Lennard-Jones potential [16]. One must sum all forces acting on each particle within the cutoff radius $R_c$, beyond which they become identically zero, since the same $R_c$ applies to all three forces. All masses are taken as equal to one. Español and Warren [12] showed that the weight functions and constants that appear in Eqs. (3) - (5) can be chosen arbitrarily, as long as they fulfill the restrictions imposed by Eq. (6), which is a consequence of the fluctuation-dissipation theorem [10 –12, 22, 27]:

$$k_B T = \sigma^2/2\gamma , \tag{6}$$

where $k_B$ is Boltzmann's constant and $T$ is the absolute temperature. For the case of $\omega^D(r_{ij})$, it is also dimensionless (see Eqs. (4) and (5)) and is defined as

$$\omega^D(r_{ij}) = \omega^R(r_{ij})^2 \quad , \tag{7}$$

which is how the fluctuation-dissipation theorem becomes operative here. The probability distribution function obeyed by these dynamics is the one corresponding to the canonical ensemble, where the total number of particles, *N*, the total volume, *V*, and *T*, are conserved. To model the polymers, we employed an additional conservative force that connects beads belonging to the same polymer [27]. The model used for such a force is a spring with constant *K* and equilibrium distance $r_{eq}$ defined by Eq. (8)

$$F_{ij}^{spring} = -K(r_{ij} - r_{eq})\hat{e}_{ij}. \tag{8}$$

Over the years, DPD simulations have been used to study polymer solutions and other soft matter systems [15 – 21, 29, 30]. Some recent examples are scaling laws that have been proposed using DPD [18, 19], which predict how liposomes behave under flow since they are an excellent medium for drug delivery [21]. The mechanical response of cancerous cells has also been modeled recently with DPD [29], which shows the versatility and usefulness of the technique. For further details and applications of DPD, the reader is referred to references [13–23], and [26–31]. Since DPD has been a very successful tool for the study of the physicochemical properties of polymer solutions, we used it here to model a series of polysiloxanes, namely polydimethylsiloxane (PDMS), polymethylhexilsiloxane (PMHS), and polydimethyl-A-1,1´, dimethylpropylnaminemethylsiloxane (P2DMPAS), in solvents that are out of theta conditions and also in theta conditions. Our primary purpose was to use the thermodynamic and structural information obtained from the DPD simulations to determine the quality of the different solvents we have modeled (toluene and water) concerning the various polysiloxanes.

**Methodology and Simulation Details**

All simulations reported in this work were performed in reduced units (marked with asterisks). The global average density of the simulation box was set equal to $\rho^* = 3$ because it has been shown that this is the lowest density for which the quadratic equation of state of DPD ($P = \rho k_B T + \alpha a \rho^2$, where *P* is the pressure, $\rho$ is the density, *a* is the interaction constant for $F^C$ in Eq. (3) and $\alpha = 0.1$, see [22]) obeys a scaling law that makes it independent of the choice of the particle-particle interaction parameters, $a_{ij}$. $R^*_c$ was taken as equal to 1, as was also the mass [27]. An additional advantage of DPD over other numerical modeling techniques is its robustness to finite size effects [26], which is a consequence of the short-range nature of its interactions; therefore, one can obtain meaningful results with relatively small systems. The dimensions of the box were a cube of a volume of 1000 with each side $L^*_x$, $L^*_y$ and $L^*_z$ of distance 10, containing 3000 beads inside. Distances were reduced with the cutoff radius, $R^*_c$, which for a coarse-graining degree equal to three water molecules per DPD particle was equal to $R_c$= 6.46 Å. Also, the reduced thermal energy is $k_BT^*$ = 1 for all of our simulations. The parameters defining the dissipative and random forces intensities were $\gamma^*$ = 4.5 and $\sigma^*$ = 3.0, respectively. The equations of motion were integrated numerically using the so-called velocity-Verlet DPD algorithm, with a time step, $\Delta t^*$, equal to 0.03 [12, 26, 27]. The simulations were run in blocks of $10^4$ time steps, with five blocks used for reaching equilibrium and fifty more for the production phase. Every DPD bead in this work had a volume equal to 90 Å$^3$, with the chemical groups per bead listed in Table 1. The conservative force intensities for interactions between particles of the same type were chosen as $a_{ij}$ = 78.0 for water, and $a_{ij}$ = 39.0 for toluene when $i = j$. This is another novelty of the present work since most reports on DPD use a coarse-graining degree equal to one rather than three, which yields interaction parameters $a_{ij}$= 25.0, which is too small to yield useful mesoscale information. Our coarse-graining degree is illustrated in Fig. 1. This soft-sphere interaction (the parameter for different types of particles, $a_{ij}$) can be mapped onto the Flory-Huggins theory through the relation [15, 27]

$$a_{ij} \approx a_{ii} + 3.27 \chi_{ij}, \tag{9}$$

where $\chi_{ij}$ was obtained from Eq. (10) [32, 33]

$$\chi_{ij} = \frac{V}{k_B T}(\delta_i - \delta_j)^2 . \tag{10}$$

Using group contribution theory, one can calculate the individual solubility parameters of each group ($\delta_i$) employing the following equation [33]:

$$\delta_i = \frac{\sum F_i}{V}, \tag{11}$$

where $F_i$ is the molar attraction constant for all the atoms that are included inside of the species $i$ and $V$ is the volume of the DPD bead. The solubility parameters of the groups that form each polymer were calculated using several molar attraction constants ($F_i$) according to either Hoy, Small or Van Krevelen data, depending on the case [29, 30, 32, 33]. For the molar attraction constant corresponding to silicon, we calculated its corresponding $F_i$ using Eq. (11) and the experimental solubility parameter for PDMS (14.9 MPa$^{1/2}$) [34]. The reader will find all the $\chi_{ij}$ and a$_{ij}$ for all the beads used here presented in Tables 2 and 3.

As described above, any polymer-solvent interactions contained in the conservative force intensities for interactions between particles can be obtained via Eq. 9. For this case, we call it ι (Greek letter nu, not to be confused with velocity)-solvent conditions. When the $\chi_{ij}$ are equal to zero, then $a_{ij} = a_{ii}$ and this condition imply that there are no interactions that are related to excluded volume interactions. This solvent condition is easily recognized as the so-called theta temperature and solvent conditions, which will be referred to as $\theta$ - solvent conditions. However, in DPD the good solvent condition has been modeled regarding $a_{ij} < a_{ii}$, as reported elsewhere [18, 20, 30]. This condition can never be obtained via ι – solvent conditions (as in this work) since the $\chi_{ij}$ obtained through the solubility parameters (Eq. (10)) is always positive. Therefore, to satisfy the solvent condition where $a_{ij} < a_{ii}$ we propose here two new solvent conditions: one which will be referred to as the ϖ – solvent condition, based on Eq. (12), and the ξ-solvent conditions, based on Eq. (13), both as shown below.

The ϖ – solvent condition uses the value of $\chi_\theta = 0.5$, which is associated with the value of the Flory interaction parameter ($\chi$) in theta-solvent conditions, to obtain a more accurate value of $a_{ij}$. The subtraction of this value from the calculated value of $\chi_{ij}$ in Eq. (12) allows us to obtain values of $a_{ij}$ that satisfy the conditions of $a_{ij} < a_{ii}$, that are typical for the DPD simulation of good solvent conditions, as stated previously [18, 20, 30]. The ξ-solvent condition uses the same concept but is limited to being applicable only for the polymer-solvent interactions ($a_{ij}$) where $a_{ij} < a_{ii}$, leaving the polymer-polymer interactions ($a_{ij,\ i \neq j}$) as Eq. (9) states, with $a_{ij} > a_{ii}$.

$$\varpi = a_{ii} + 3.27(\chi_{ij} - 0.5) \qquad (12)$$

$$\xi = a_{ii} + 3.27\delta_{ij}$$
$$\delta_{ij} = \begin{cases} \chi_{ij} & \text{for all polymer} - \text{polymer} \quad a_{ij, i \neq 1} \\ \chi_{ij} - 0.5 & \text{for all polymer} - \text{solvent} \quad a_{ij, i = 1} \end{cases} \qquad (13)$$

We used five polymer molecules in the box with polymerization degree each equal to 102 DPD beads, while the rest of DPD beads in the box were solvent molecules. The segmentation of the polysiloxanes studied here is shown in Fig. 2, where their coarse-grained model division into different DPD beads is shown. Figure 2 B shows the mapping of the PDMS into the DPD beads, Fig. 2C shows the mapping of PMHS, while Fig. 2D presents the DPD mapping of P2DMPAS. The mapping of the solvents is shown in Fig. 2A. Once the polymers were segmented into several DPD beads joined by springs, we proceeded to obtain the solubility parameters for each DPD bead, corresponding to each chemical group, and with the solubility parameter obtained from Eq. (11), we calculated the Flory interaction parameter, $\chi_{ij}$, for all the interactions between each DPD bead present in the system of study. The resulting $\chi_{ij}$'s describe the interaction between each DPD bead; for example, in the case of PDMS $\chi_{12}$ is the interaction between the solvent DPD bead (DPD bead 1) and the dimethylsiloxane bead (DPD bead 2).

**Results and Discussion**

**PDMS as a case study: density profiles and radial distribution functions**

Our first approach to understanding how DPD simulation can be applied to real polymer-solvent mixtures was studying a well-known polymer, namely PDMS. As is well known, PDMS has been reported to have a good solvent interaction with toluene and a poor solvent interaction with water; this behavior was reported based on both the solubility parameters and $A_2$ values [2, 4, 8, 34]. On the basis of these experimental results we simulated PDMS in both solvents varying the solvent condition ($\theta$, $\iota$, $\varpi$ and $\xi$). We focused on three structural properties: density profile, radial distribution function, and conformational snapshots.

In Fig. 3 the density profiles of the solvent and main chain beads for both solvents (water and toluene) for all solvent conditions used are depicted. For both solvents the $\theta$ solvent condition density profiles (Figs. 3A and 3C) were similar, maintaining the density of both beads almost constant through the $z$-axis. This behavior is what is expected for a solvent in theta conditions where there is no preferential interaction between the particles in the box, even if their respective conservative force parameters $a_{ii}$ (78 and 38, respectively) are different. When the solvent condition was changed to $\iota$ (the original form of $a_{ij}$ proposed by Groot et al. [22]), we observed that in water the solvent beads showed a very repulsive nature towards the main chain beads (Figs. 3A and 3C) while for toluene the repulsion between beads was less intense. If toluene were a good solvent, no repulsion would be observed. For this reason, we investigated the other two solvent conditions ($\varpi$ and $\xi$) which are depicted in Figs. 3B and 3D. For water (Fig. 3B) the $\varpi$ condition exhibited some repulsion between solvent and polymer beads through the z-axis while for the $\xi$ condition there was a large repulsion between polymer and solvent molecules, similar to that observed for the $\iota$ condition. By contrast, in toluene (Fig. 3D) both solvent conditions, $\varpi$ and $\xi$, behaved in a similar way as for the $\theta$ conditions, meaning that the polymer chains were well dispersed through the box.

Following the density profile analysis, we studied the structural properties of the fluid through the study of the radial distribution function (RDF) corresponding to the interactions (contained in the $a_{ij}$ term) between the main chains and the solvent ($a_{12}$), and the interactions between the main chains and the end groups ($a_{23}$). These interactions correspond to those of the main chain (dimethylsiloxane group = DPD bead 2, see Fig. 2) with their terminal ends (trimethyl siloxane groups = DPD bead 3, see Fig. 2) and the solvent (DPD bead 1) for both solvents and for all conditions ($\theta$, $\iota$, $\varpi$ and $\xi$). Figure 4B shows the RDFs corresponding to toluene for every non-theta solvent condition discussed here ($\iota$, $\varpi$ and $\xi$); they were all larger than the ones corresponding to water. This means that it is more probable to find a toluene DPD bead than a water DPD bead near a PDMS bead under the same solvent conditions. Also, we observed that for the case of $\theta$-solvent conditions both RDFs (water -Fig 4A- and toluene -Fig 4B-) were equal and had the largest intensity values. The interactions between end group beads and main chain beads showed the opposite trend, where the RDFs in water ($a_{23\iota}$, $a_{23\varpi}$ and $a_{23\xi}$ Fig 4A) exhibited a much larger intensity for every non-theta condition than those in toluene ($a_{23\iota}$, $a_{23\varpi}$ and $a_{23\xi}$ Fig. 4B). We observe that the differences between water RDFs and toluene RDFs were three times greater at their respective peak for every curve depicted in the figure, at the particularly favored separation distance (around $r^*=0.95$). This implies that it is much more probable to find a compact conformation for the PDMS in water that in toluene, and for the second favored distance ($r^*=1.5$) this difference was also maintained.

Finally, the snapshots of the simulations (Fig. 5) revealed that the PDMS chains were uniformly distributed in the simulation box for all the solvent conditions in toluene, implying that the main chain is always solvated, while in water the opposite was the case when the solvent conditions were different from θ-solvent conditions. This result is in agreement with the experimental data [34] and means that toluene is a good solvent for this polymer.

**PMHS: a polymer where the lateral group influences the conformation of the polymer almost as much as the main chain.**

In the case of PMHS very little information about this polymer is available in the literature. It was reported that toluene also was a good solvent for this polymer, with values of $A_2$ of $10^{-4}$ mol/mlg$^2$ [3]. This value of $A_2$ also indicates that, even if toluene is a good solvent for this polymer, the value was close to zero, which is the theoretical value for a theta solvent. This result implies that toluene is close to a theta solvent for PMHS. Therefore, we considered it important to model this complex polymer to determine how it behaves when one varies the solvent and solvent conditions.

An analysis of the density profiles, which are depicted in Fig. 6, reveals that even under θ solvent conditions water exhibited some degree of repulsion for the solvent beads against the polymer (Fig. 6A) which could be due to the lateral groups of this polymer. This behavior was not observed for toluene (Fig. 6C) where the density profile was nearly constant through the *z*-axis. For the ι condition in water, PMHS showed a less intense repulsion than PDMS for the solvent beads. In the case of toluene, the density profile of the solvent beads was also constant through the *z*-axis. As shown in Fig. 6B, for the ϖ condition in water, the repulsion was more pronounced than the one observed for the ι condition; in the case of toluene, the behavior was similar to what was observed for the θ-condition. Finally, in the same figure, when the ξ-condition was analyzed, we found that in the case of water the repulsion of the solvent beads was equal to that observed for the ι-condition, while in toluene the behavior of the density profile was similar to the one observed for the θ-condition.

The RDFs corresponding to the interactions of the main chain with the solvent ($a_{12}$) and the lateral groups ($a_{23}$) and the interaction between the solvent and the lateral groups ($a_{13}$) are presented in Fig. 7. Figure 7 shows that the main chain-solvent interactions corresponding to toluene (Fig. 7B) were greater than those for the water case (Fig. 7A), similarly to what was observed for PDMS (Fig. 4) and were about equal to the RDFs of solvent qualities ι, ϖ, and ξ. The same figure displays the same behavior for both solvents in the case of the interactions between the lateral groups and the solvent ($a_{13}$). For the interactions between the lateral group with the main chain ($a_{23}$), depicted in Fig. 7A in the case of water, the RDF corresponding to the case under θ solvent conditions is the smallest of the four solvent conditions. The other three conditions (ι, ϖ, ξ) have smaller values of their RDFs at its first maximum, which means that it is more probable that the beads along the main chain find a lateral group in water. For toluene (Fig. 7B) the RDFs were very similar in the four solvent conditions, implying that the behavior in toluene out of *θ* – conditions is very similar to the one under *θ* – conditions, which is in agreement with the low value of $A_2$ reported for toluene elsewhere [3].

Finally, from the conformational snapshots of all of the solvent conditions, shown in Fig. 8, one finds that PMHS exhibited a better dispersion of the chains in the simulation box than PDMS (Fig. 5), in the case of water for every solvent condition. In the case of toluene, good dispersion of the PMHS polymer chains was also observed under every solvent condition revealing that toluene is a good solvent since no agglomeration of chains was observed.

**How a polar group influences the conformation of P2DMPAS.**

In a recent study, the solubility of a PDMS based random copolymer was modified by functionalization with a polar group [6]. This is an important phenomenon that should be addressed to understand how the functionalization changes the solubility and the interactions that take place between the solvent and the polymer. Following this reasoning we considered it relevant to simulate a copolymer which is mostly PDMS but with some amino lateral groups, looking for a change in the conformation of the polymer compared to pure PDMS and implying a possible change in its solubility in certain solvents. The polymer, called P2DMPAS, was simulated in the same two solvents (water and toluene) as the other two polymers modeled in this study (PDMS and PMHS), varying the solvent conditions (θ, ι, ϖ and ξ) similarly.

First, the density profiles of the solvent and main chain beads (for this case the main chain was composed of two types of beads: 2 = dimethylsiloxane groups and 3 = methylsiloxane groups) are depicted in Fig. 9. Figure 9A shows that in water the *θ* – solvent condition displayed a little repulsion of the solvent beads towards the polymer beads. In the case of the ι-condition, the repulsion became evident, suggesting the formation of agglomerates of polymer chains. In Fig. 9B the ϖ-condition and ξ-condition exhibited the same repulsion observed under the ι-condition for water; this is the same behavior as observed for PDMS. In the case of toluene, in Fig. 9C, the θ-condition showed a constant behavior of the solvent beads through the $z$-axis, while the ι-condition exhibited a small repulsion of the solvent beads towards the main chain. For the ϖ-condition, depicted in Fig. 9D, the density profile of the solvent beads tended to be constant through the $z$-axis, as in the θ-condition. For the case of the ξ-condition a slight repulsion was found, similar to the one observed under the ν condition.

To corroborate what was found in the density profiles of the polymer and solvents, RDF analyses of the main chain-solvent ($a_{12}$, $a_{13}$), main chain-polar group ($a_{25}$, $a_{35}$), main chain-main chain ($a_{23}$) and polar group-solvent ($a_{15}$) interactions were made. They are displayed in Fig. 10. As Figs. 10A and B show, the main chain-solvent interactions were similar to those shown in Fig. 4A for PDMS, where the RDFs corresponding to toluene (Fig. 10B) was larger than the ones for water (Fig. 10A) for every non-theta solvent condition. In the case of the interaction between the polar groups and the solvent ($a_{15}$, $b_{15}$), depicted in Fig. 10, the RDFs were similar to those discussed above. For water, (Fig. 10A) the magnitude of these three RDFs tended to increase as the relative bead to bead distance $r^*$ increased, but all were smaller than the ones for toluene (Fig. 10 B). This means that toluene interacted more with P2DMPAS than the latter did with water.

For the main chain-main chain interactions ($a_{23}$) the opposite behavior was observed, as shown in Fig. 10, with all the non-theta solvent condition having essentially the same RDFs. Also, their magnitude is greater than those for the main chain-solvents described above. Concerning the interactions between the main chains and the polar groups, Fig. 10 show the RDFs of water were larger than those for toluene. In the case of the dimethylsiloxane bead-amino group interaction ($a_{25}$) in toluene (Fig. 10B) it is worth noting that the first and second most frequent correlation distances corresponded to peaks in the RDF of similar magnitude, being the largest for the ι-condition. This implies that it is almost equally probable to find a polar group as a first or second neighbor of the polydimethyl group. Overall, P2DMPAS showed behavior similar to PDMS for both solvents studied here.

The configurational snapshots of P2DMPAS are presented in Fig. 11A, which shows the agglomeration of the chains in water, similar to what was observed for PDMS (see Fig. 5). The number of polar groups per chain (less than 5) was not sufficient to play an important role in the conformation as the main chain was still governing the solubility of the polymer because most of it was made up of PDMS beads. This also explains, why in toluene, the chains exhibited good dispersion in the entire box (Fig. 11B).

**Conclusions**

We have found that DPD is a useful tool to study the behavior of complex polysiloxanes under various solvent conditions; it revealed the trend of the siloxane bond to interact preferentially with siloxane bonds in other chains, rather than with other types of beads. Also, DPD was helpful for the prediction of solvent conditions given the structure of the polymer as we observed that it replicated the experimental behavior observed for a common polymer such as PDMS. In the case of a more complex polymer (such as P2DMPAS), it remains to be determined if the polar group is charged as it interacts with the polar solvents. A further study is also necessary to determine how the solubility of P2DMPAS changes as the number of polar groups increases in its main chain. It is noted this simulation technique offers a good opportunity to study theta conditions since they are very difficult to obtain experimentally. Finally, we conclude that the regular solvent condition ($\iota$) is very well suited for poor solvents; however, in the case of good solvents, the $\varpi$ and $\xi$ are considerably better since they appear to be near the $\theta$-solvent condition.

**Acknowledgments**

We would like to thank Adan Bazan for technical support. The Laboratorio Nacional de Caracterización de Propiedades Fisicoquímicas y Estructura Molecular Supercómputo Universidad de Guanajuato is gratefully acknowledged for the generous allocation of computer time for the simulations. AGG acknowledges the computer resources and support provided by the Laboratorio Nacional de Supercómputo del Sureste de México, CONACYT network of national laboratories.

**Figure Captions**

**Figure 1.** Schematic representation of the conservative force interaction term in DPD, see Eq. (3).

**Figure 2.** Schematic mapping of the polysiloxanes into DPD polymers. A) Solvent DPD bead for the solvents employed; B) Mapping of PDMS into its corresponding DPD beads; C) Mapping of PMHS into its corresponding DPD beads; D) Mapping of P2DMPAS into its corresponding DPD beads.

**Figure 3.** Comparison of the density profiles along the z-direction $\rho(z^*)$ for the solvent beads (Bead 1) and main polymer chain bead (Bead 2) in every solvent condition simulated ($\theta, \iota, \varpi, \xi$) for PDMS. A) $\theta, \iota$ solvent conditions in water, B) $\varpi, \xi$ solvent conditions in water, C) $\theta, \iota$ solvent conditions in toluene, D) $\varpi, \xi$ solvent conditions in toluene. The polymer beads corresponding to the end groups are not shown for clarity.

**Figure 4.** Radial distribution functions (RDF) corresponding to the main chain (dimethylsiloxane, DPD bead 2) interactions ($a_{12}$, $a_{23}$) for PDMS: A) water; B) toluene. The notation for this figure is as follows: $a_{ij}$ refers to the term of the conservative force interactions between two different DPD particles (*i* and *j*), see the first expression in Eq. (3). The Greek letters refer to the solvent condition of the solvent used.

**Figure 5.** Configurational snapshots of the PDMS in A) Water, and B) Toluene, varying the solvent conditions. The solvent beads are omitted for clarity of the polymer conformations.

**Figure 6.** Comparison of the density profiles along the z-direction, $\rho(z^*)$, for the solvent beads (Bead 1) and main chain bead (Bead 2) in every solvent condition simulated ($\theta, \iota, \varpi, \xi$) for PMHS. A) $\theta, \iota$ solvent conditions in water, B) $\varpi, \xi$ solvent conditions in water, C) $\theta, \iota$ solvent conditions in toluene, D) $\varpi, \xi$ solvent conditions in toluene. The density profile of the polymer beads corresponding to the lateral and end groups are not shown for clarity.

**Figure 7.** RDF's corresponding to the main chain (methylsiloxane groups= DPD bead 2), lateral groups (ethylene groups= DPD bead 3) and solvent (DPD bead 1) interactions ($a_{12}$, $a_{13}$, $a_{23}$) for PMHS, see Figure 1.A) water; B) toluene. The notation for this figure is as follows: $a_{ij}$ refers to the term of the conservative force interaction (see first expression in Eq. (3)) between two different DPD particles (*i* and *j*). The Greek letters refer to the solvent condition of the solvent used.

**Figure 8.** Configurational snapshots of the PMHS in A) Water and B) Toluene varying the solvent conditions ($\theta$, $\iota$, $\varpi$ and $\xi$). The solvent beads are omitted for clarity to permit appreciation of the polymer conformations.

**Figure 9.** Comparison of the density profiles $\rho(z^*)$ for the solvent beads (Bead 1) and main chain beads (Beads 2 and 3) in all solvent condition simulated ($\theta$, $\iota$, $\varpi$, $\xi$) for P2DMPAS. A) $\theta$, $\iota$ solvent conditions in water, B) $\varpi$, $\xi$ solvent conditions in water, C) $\theta$, $\iota$ solvent conditions in toluene, D) $\varpi$, $\xi$ solvent conditions in toluene. The polymer beads corresponding to the lateral and end groups are not shown for simplicity.

**Figure 10.** RDFs corresponding to the main chain (DPD beads 2 {dimethylsiloxane} and 3{methylsiloxane}), the amino group (DPD bead 5) and solvent (DPD bead 1) interactions ($a_{12}$, $a_{13}$, $a_{15}$, $a_{23}$, $a_{25}$, $a_{35}$) for P2DMPAS. A) water; B) toluene. The notation for this figure is as follows: $a_{ij}$ refers to the term of the conservative force interaction between two different DPD particles (*i* and *j*, see first expression in Eq. (3)). The Greek letters refer to the solvent condition of the solvent used.

**Figure 11.** Configurational snapshots of the polymer P2DMPAS in A) Water and B) Toluene varying the solvent conditions ($\theta$, $\iota$, $\varpi$ and $\xi$). The solvent beads are omitted for clarity.

**Table Captions**

*Table 1. Polymer segmentation in DPD beads*

| Polymer/Solvent | Chemical Group DPD Bead | DPD Beads |
|---|---|---|
| **Toluene** | Toluene ($C_6H_5CH_3$) | 1 |
| **Water** | Water ($H_2O$) | 1 |
| **PDMS** | Dimethylsiloxane (-$2CH_2SiO$-) | 2 |
| | Trimethylsiloxane ($3CH_3SiO$-) | 3 |
| **PMHS** | Methylsiloxane (-$CH_3SiO$-) | 2 |
| | Ethylene (-$CH_2CH_2$-) | 3 |
| | Methylmethylene (-$CH_2CH_3$) | 4 |
| | Trimethylsiloxane ($3CH_3SiO$-) | 5 |
| **P2DMPAS** | Dimethylsiloxane (-$2CH_2SiO$-) | 2 |
| | Methylsiloxane (-$CH_3SiO$-) | 3 |
| | Ethylene (-$CH_2CH_2$-) | 4 |
| | Dimethylaminomethylene ($2CH_3NCH_2$-) | 5 |
| | Trimethylsiloxane ($3CH_3SiO$-) | 6 |

*Table 2. Solubility parameter($\chi_{ij}$) for each DPD bead varying the solvent.*

| $\chi_{ij}$ of the DPD beads | | | | | | |
|---|---|---|---|---|---|---|
| | ι | | ϖ | | ξ | |
| **PDMS** | Toluene | Water | Toluene | Water | Toluene | Water |
| $\chi_{12}$ | 0.44 | 25.35 | -0.06 | 24.85 | -0.06 | 24.85 |
| $\chi_{13}$ | 0.79 | 27.08 | 0.29 | 26.58 | 0.29 | 26.58 |
| $\chi_{23}$ | 0.04 | 0.04 | -0.46 | -0.46 | 0.04 | 0.04 |

| PMHS | Toluene | Water | Toluene | Water | Toluene | Water |
|---|---|---|---|---|---|---|
| $\chi_{12}$ | 0.31 | 7.57 | -0.19 | 7.07 | -0.19 | 7.07 |
| $\chi_{13}$ | 0.07 | 0.00 | -0.43 | -0.50 | -0.43 | -0.50 |
| $\chi_{14}$ | 0.64 | 0.00 | 0.14 | -0.50 | 0.14 | -0.50 |
| $\chi_{15}$ | 0.79 | 41.33 | 0.29 | 40.83 | 0.29 | 40.83 |
| $\chi_{23}$ | 0.03 | 0.03 | -0.47 | -0.47 | 0.03 | 0.03 |
| $\chi_{24}$ | 0.03 | 0.03 | -0.47 | -0.47 | 0.03 | 0.03 |
| $\chi_{25}$ | 0.05 | 0.05 | -0.45 | -0.45 | 0.05 | 0.05 |
| $\chi_{34}$ | 0.09 | 0.09 | -0.41 | -0.41 | 0.09 | 0.09 |
| $\chi_{35}$ | 0.12 | 0.12 | -0.38 | -0.38 | 0.12 | 0.12 |
| $\chi_{45}$ | 0.00 | 0.00 | -0.50 | -0.50 | 0.00 | 0.00 |

| P2DMPAS | Toluene | Water | Toluene | Water | Toluene | Water |
|---|---|---|---|---|---|---|
| $\chi_{12}$ | 0.44 | 7.82 | -0.06 | 7.32 | -0.06 | 7.32 |
| $\chi_{13}$ | 0.31 | 7.57 | -0.19 | 7.07 | -0.19 | 7.07 |
| $\chi_{14}$ | 0.07 | 6.94 | -0.43 | 6.44 | -0.43 | 6.44 |
| $\chi_{15}$ | 0.69 | 8.20 | 0.19 | 7.70 | 0.19 | 7.70 |
| $\chi_{16}$ | 0.79 | 8.35 | 0.29 | 7.85 | 0.29 | 7.85 |
| $\chi_{23}$ | 0.01 | 0.01 | -0.49 | -0.49 | 0.01 | 0.01 |
| $\chi_{24}$ | 0.11 | 0.11 | -0.39 | -0.39 | 0.11 | 0.11 |
| $\chi_{25}$ | 0.02 | 0.02 | -0.48 | -0.48 | 0.02 | 0.02 |
| $\chi_{26}$ | 0.04 | 0.04 | -0.46 | -0.46 | 0.04 | 0.04 |
| $\chi_{34}$ | 0.03 | 0.03 | -0.47 | -0.47 | 0.03 | 0.03 |

*Table 3. $a_{ij}$ for each DPD bead in every solvent for all solvent conditions*

| $a_{ij}$ of the DPD beads | | | | | | |
|---|---|---|---|---|---|---|
| | $\iota$ | | $\varpi$ | | $\xi$ | |
| PDMS | Toluene | Water | Toluene | Water | Toluene | Water |
| $a_{12}$ | 40.28 | 160.82 | 38.65 | 159.18 | 38.65 | 159.18 |
| $a_{13}$ | 41.44 | 166.49 | 39.81 | 164.85 | 39.81 | 164.85 |
| $a_{23}$ | 38.97 | 78.04 | 37.33 | 76.40 | 38.97 | 78.04 |
| **PMHS** | Toluene | Water | Toluene | Water | Toluene | Water |
| $a_{12}$ | 39.85 | 102.67 | 38.22 | 101.04 | 38.22 | 101.04 |
| $a_{13}$ | 39.08 | 77.92 | 37.45 | 76.28 | 37.45 | 76.28 |
| $a_{14}$ | 40.95 | 77.92 | 39.31 | 76.28 | 39.31 | 76.28 |
| $a_{15}$ | 41.44 | 213.08 | 39.81 | 211.45 | 39.81 | 211.45 |
| $a_{23}$ | 38.96 | 78.03 | 37.32 | 76.40 | 38.96 | 78.03 |
| $a_{24}$ | 38.93 | 78.00 | 37.30 | 76.37 | 38.93 | 78.00 |
| $a_{25}$ | 39.01 | 78.08 | 37.37 | 76.44 | 39.01 | 78.08 |
| $a_{34}$ | 39.13 | 78.20 | 37.49 | 76.56 | 39.13 | 78.20 |
| $a_{35}$ | 39.23 | 78.30 | 37.59 | 76.67 | 39.23 | 78.30 |
| $a_{45}$ | 38.86 | 77.93 | 37.22 | 76.29 | 38.86 | 77.93 |
| **P2DMPAS** | Toluene | Water | Toluene | Water | Toluene | Water |
| $a_{12}$ | 40.28 | 103.49 | 38.65 | 101.85 | 38.65 | 101.85 |
| $a_{13}$ | 39.85 | 102.67 | 38.22 | 101.04 | 38.22 | 101.04 |
| $a_{14}$ | 39.08 | 100.61 | 37.45 | 98.98 | 37.45 | 98.98 |
| $a_{15}$ | 41.09 | 104.74 | 39.45 | 103.11 | 39.45 | 103.11 |
| $a_{16}$ | 41.44 | 105.24 | 39.81 | 103.60 | 39.81 | 103.60 |
| $a_{23}$ | 38.87 | 77.95 | 37.24 | 76.31 | 38.87 | 77.95 |
| $a_{24}$ | 39.21 | 78.28 | 37.57 | 76.65 | 39.21 | 78.28 |
| $a_{25}$ | 38.91 | 77.98 | 37.28 | 76.35 | 38.91 | 77.98 |
| $a_{26}$ | 38.97 | 78.04 | 37.33 | 76.40 | 38.97 | 78.04 |
| $a_{34}$ | 38.96 | 78.03 | 37.32 | 76.40 | 38.96 | 78.03 |
| $a_{35}$ | 38.95 | 78.02 | 37.32 | 76.39 | 38.95 | 78.02 |
| $a_{36}$ | 39.01 | 78.08 | 37.37 | 76.44 | 39.01 | 78.08 |
| $a_{45}$ | 39.16 | 78.23 | 37.52 | 76.59 | 39.16 | 78.23 |
| $a_{46}$ | 39.23 | 78.30 | 37.59 | 76.67 | 39.23 | 78.30 |
| $a_{56}$ | 38.85 | 77.92 | 37.21 | 76.28 | 38.85 | 77.92 |

Figure 1

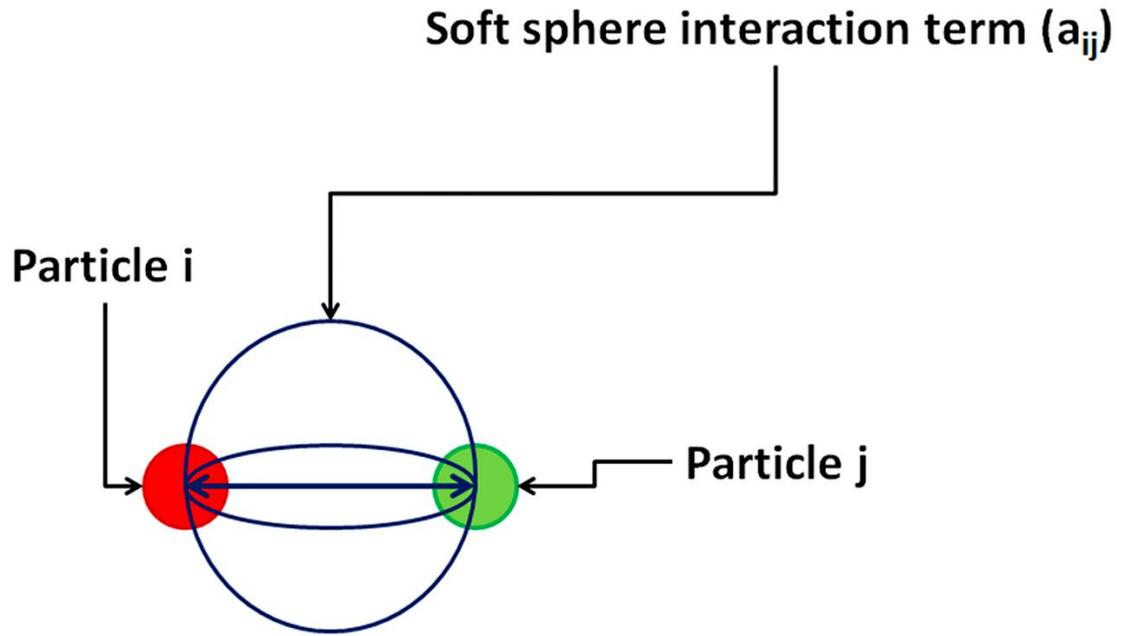

Figure 2

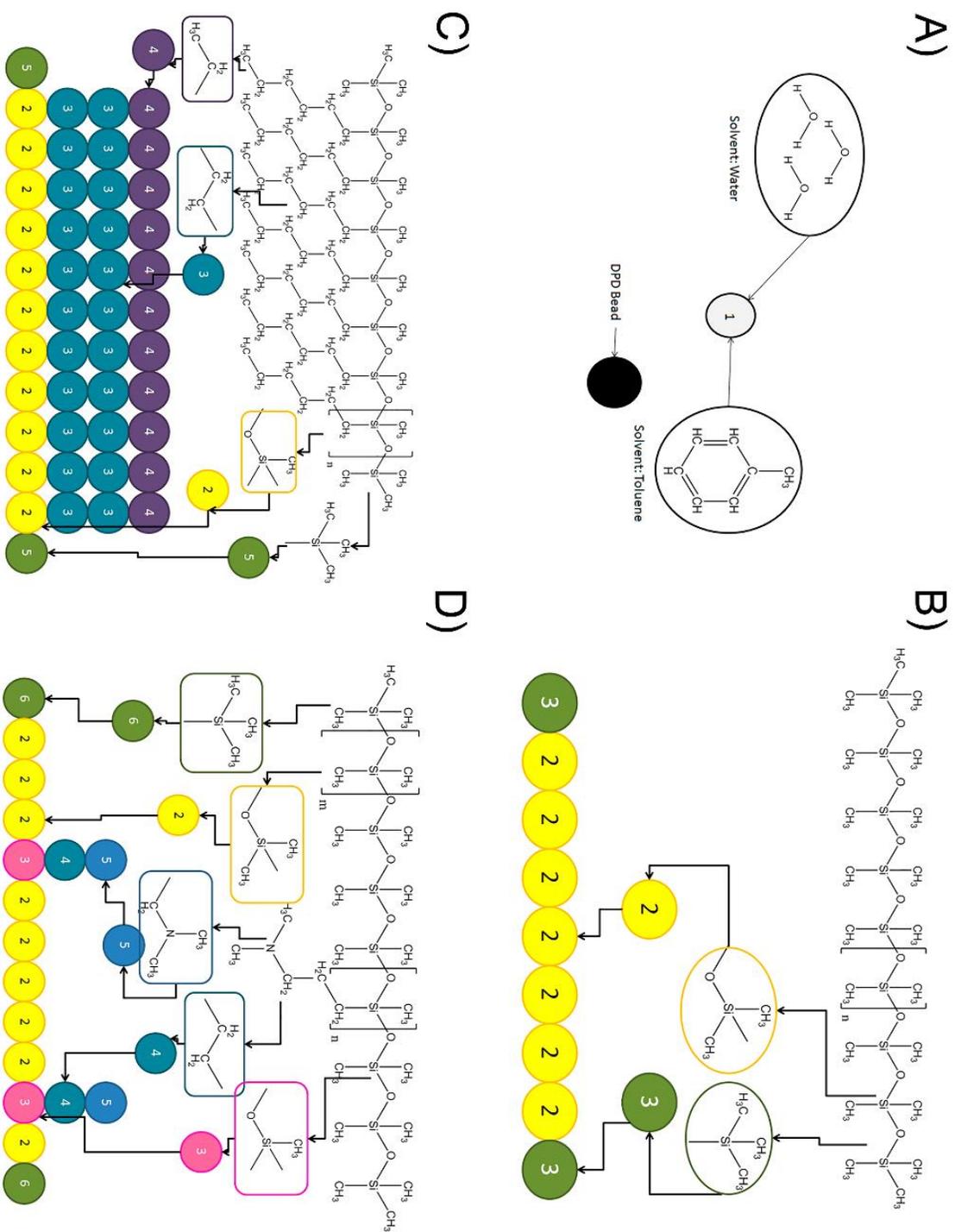

Figure 3

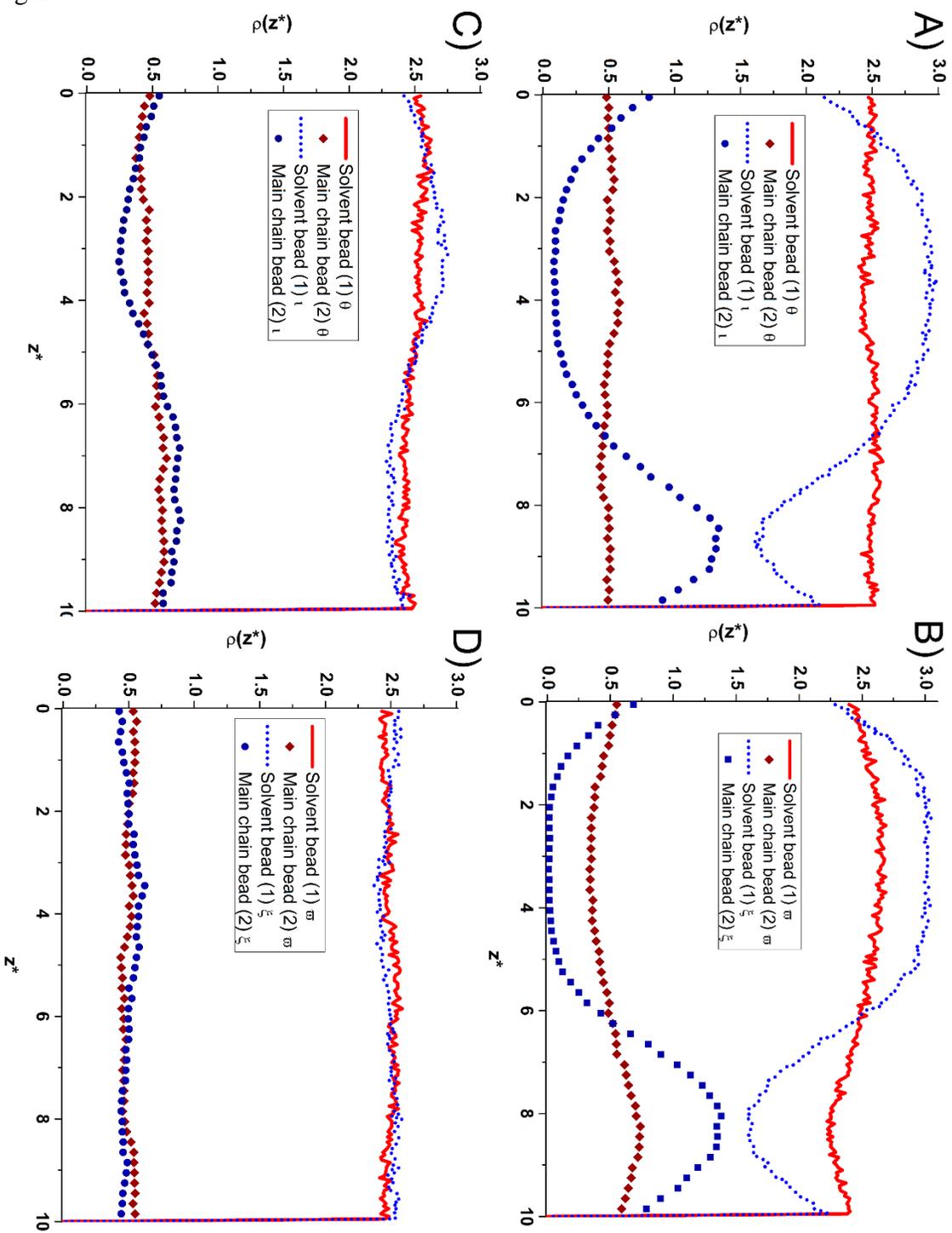

Figure 4

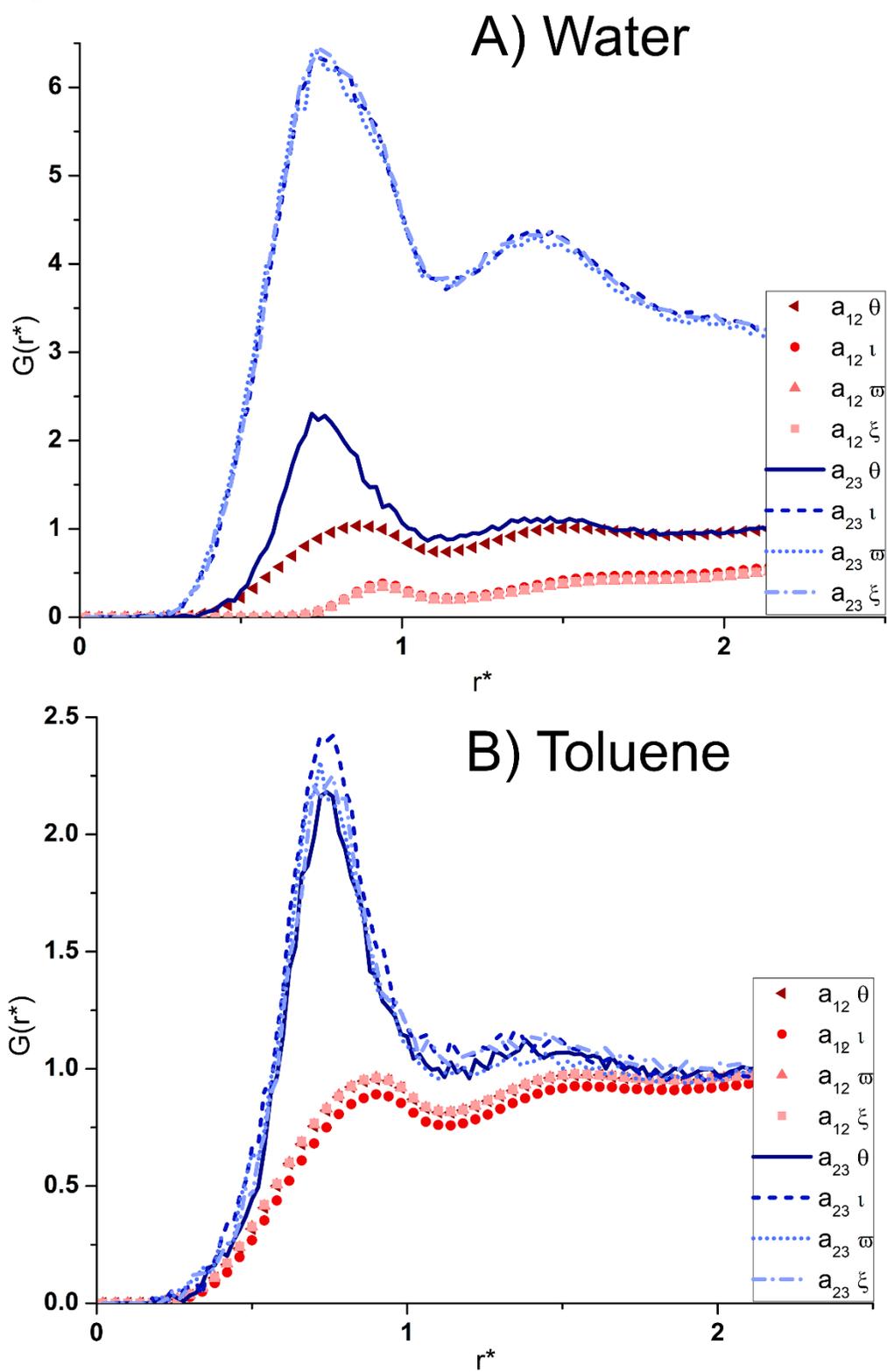

Figure 5

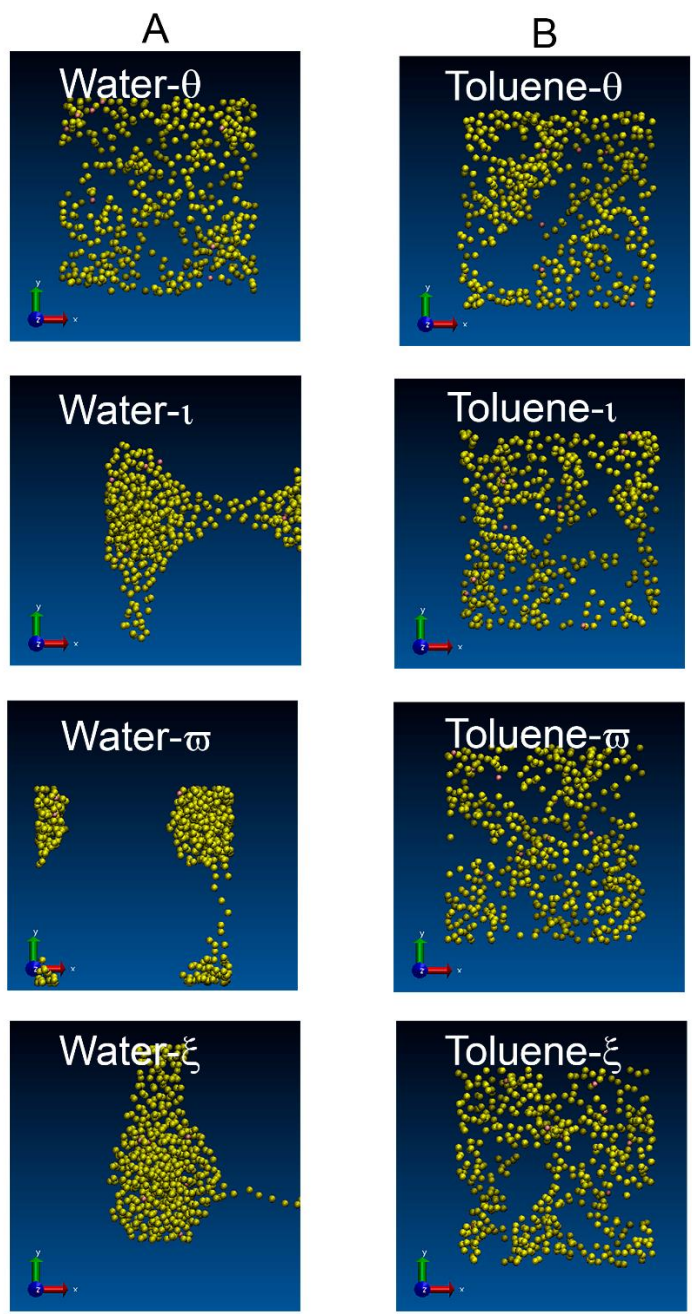

Figure 6

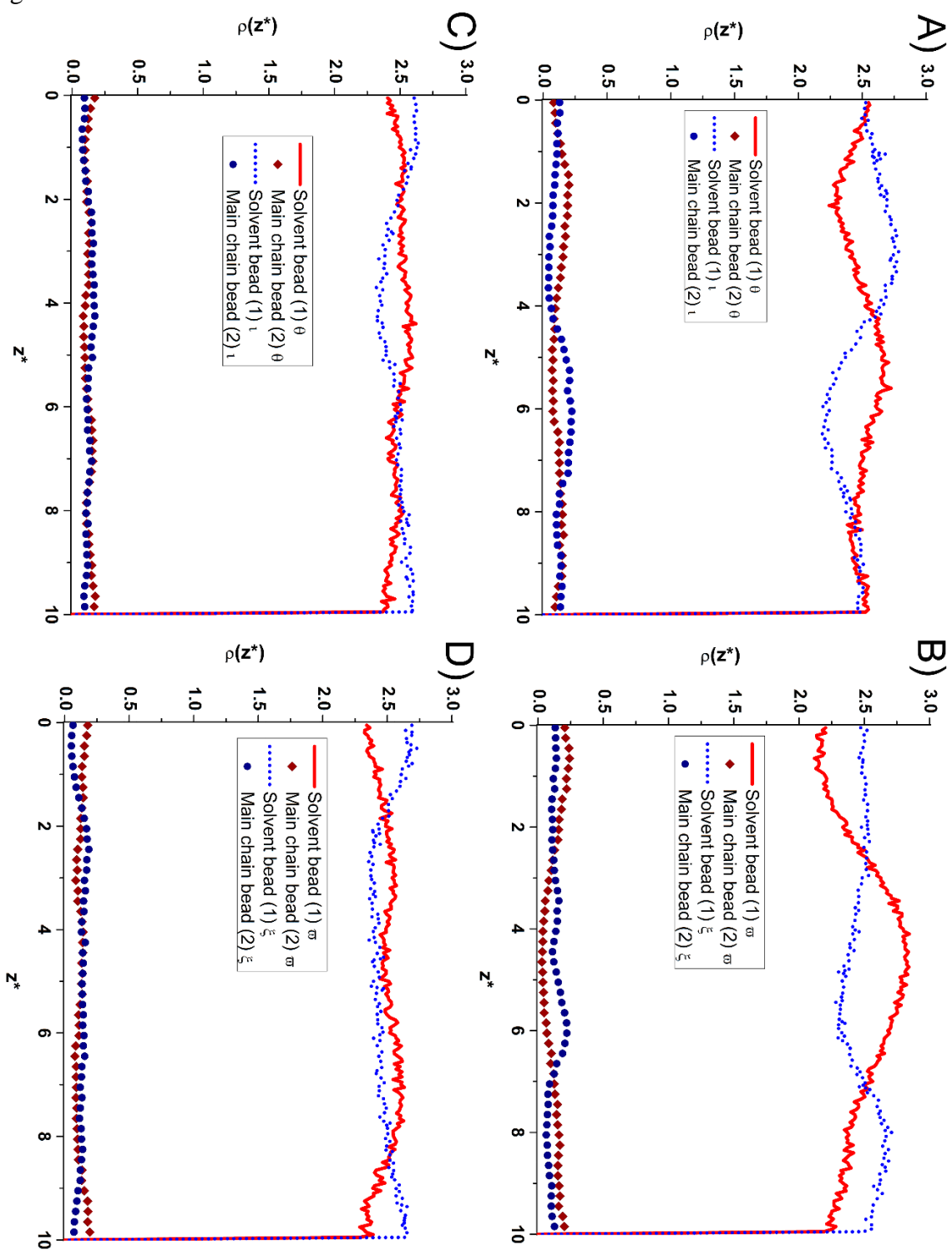

Figure 7

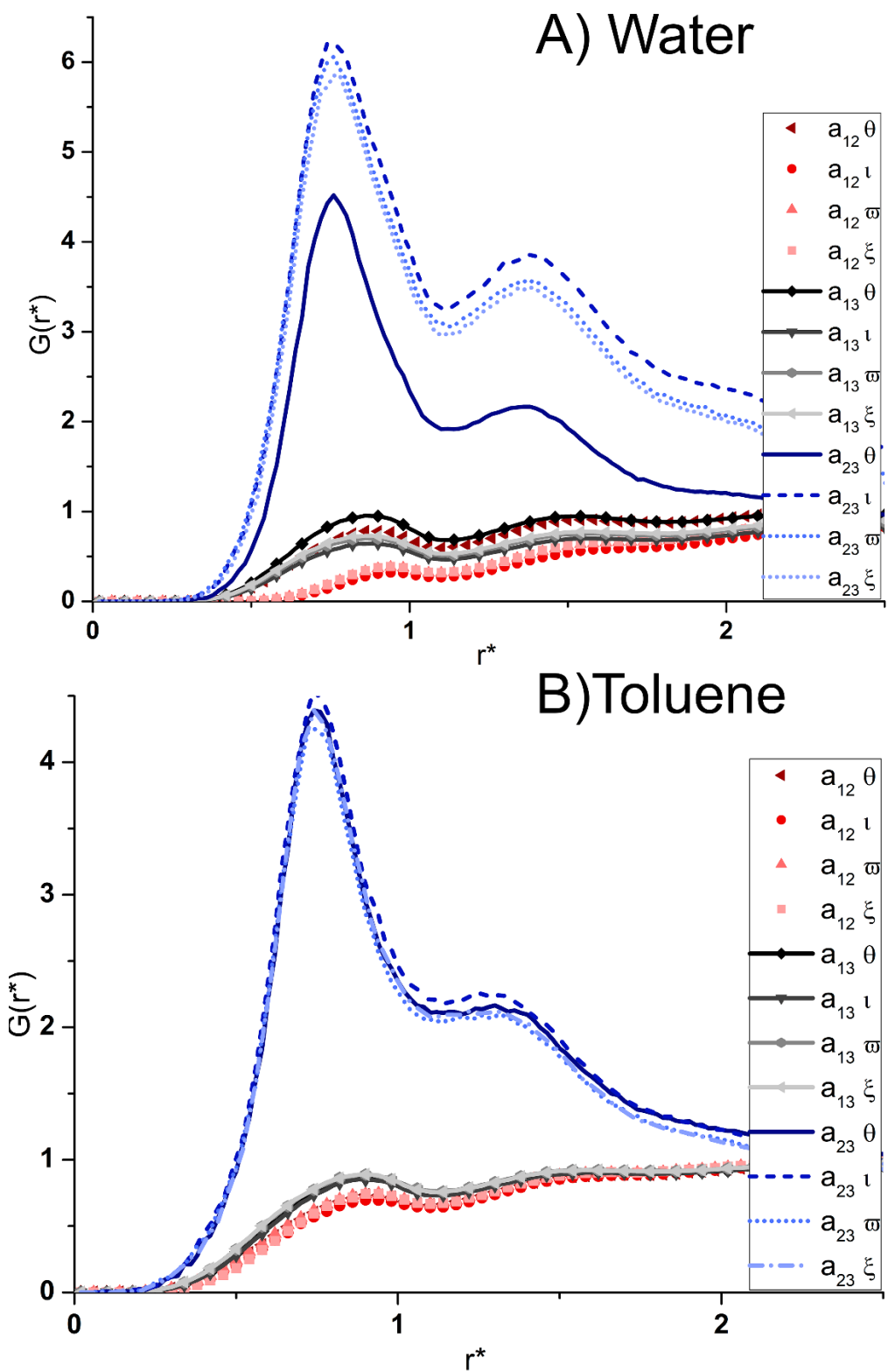

Figure 8

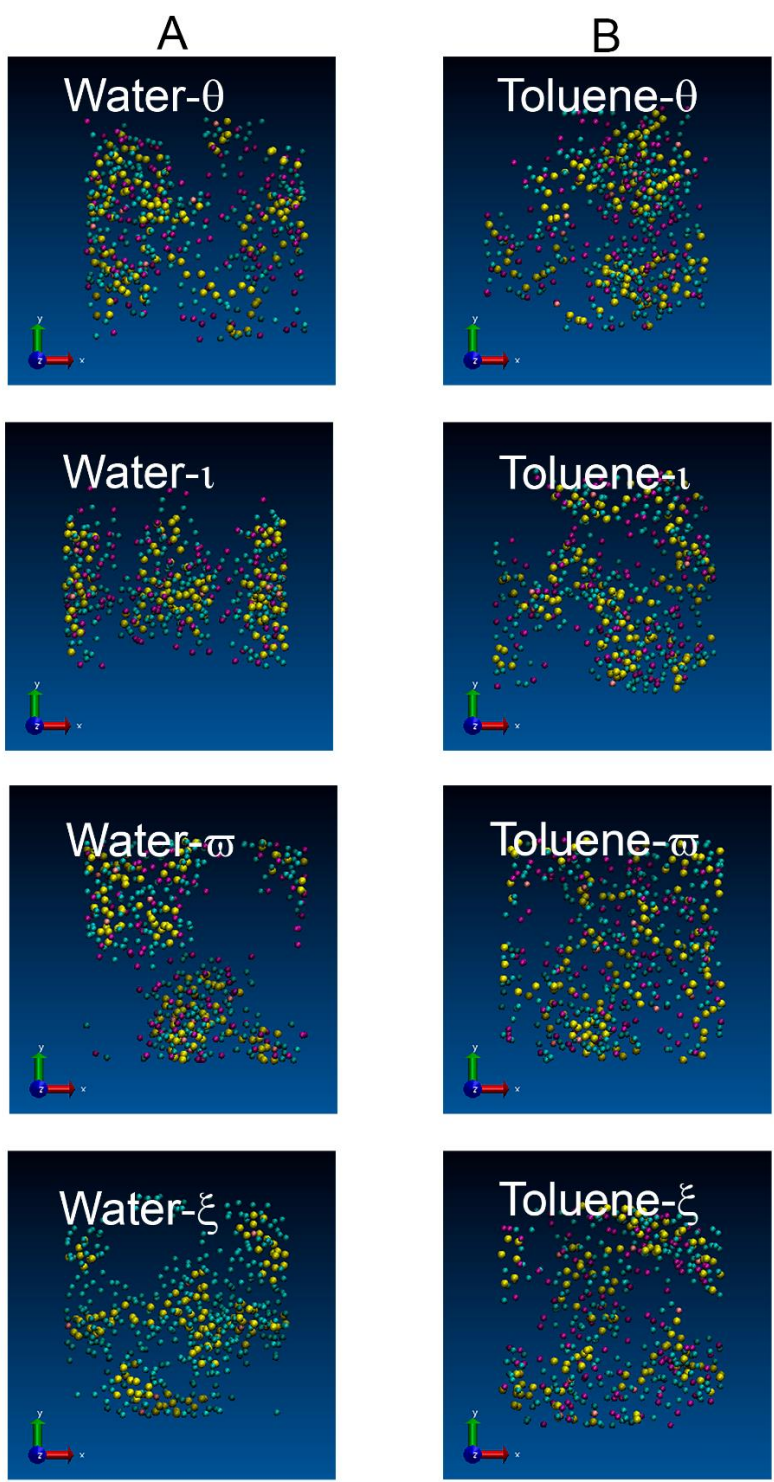

Figure 9

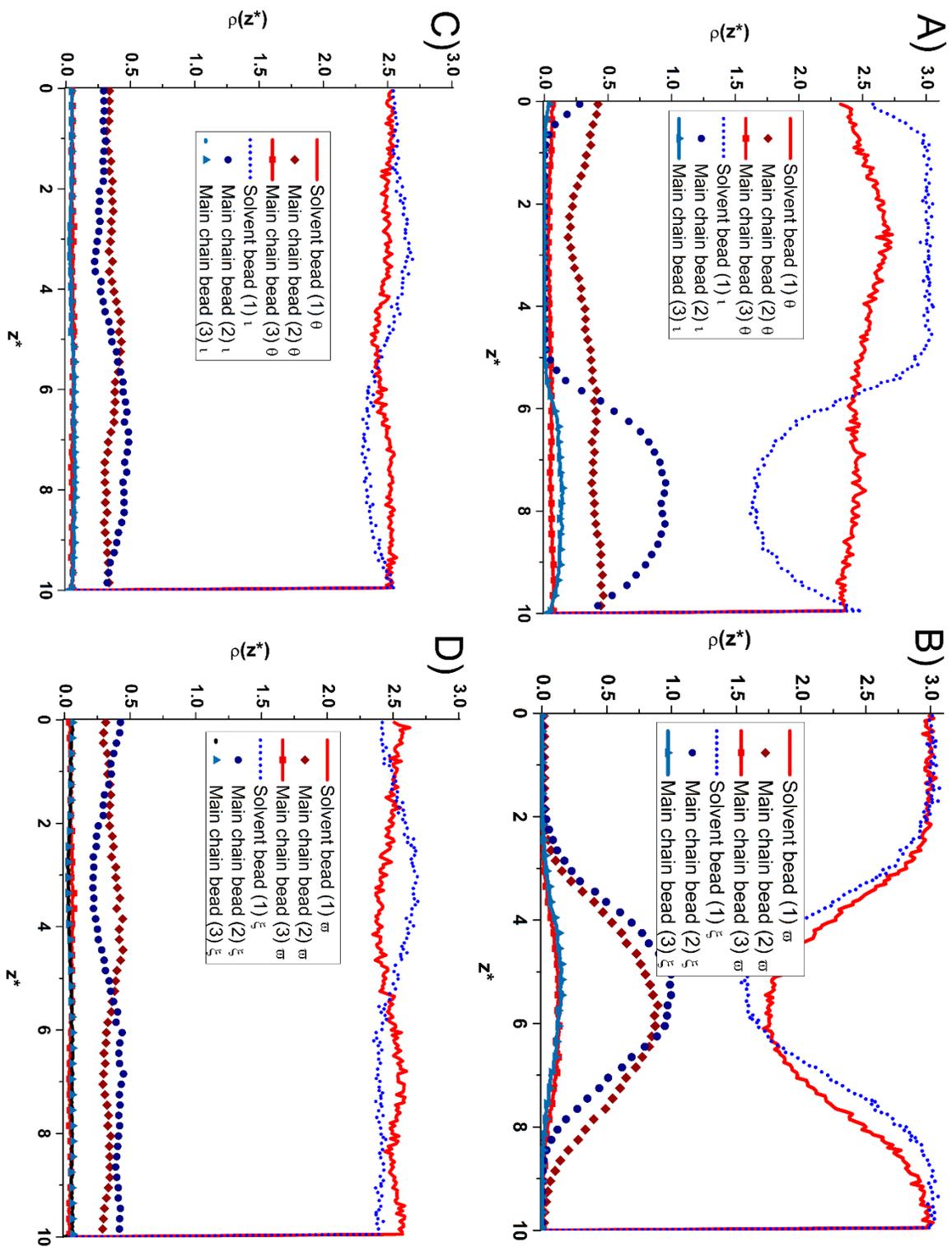

Figure 10

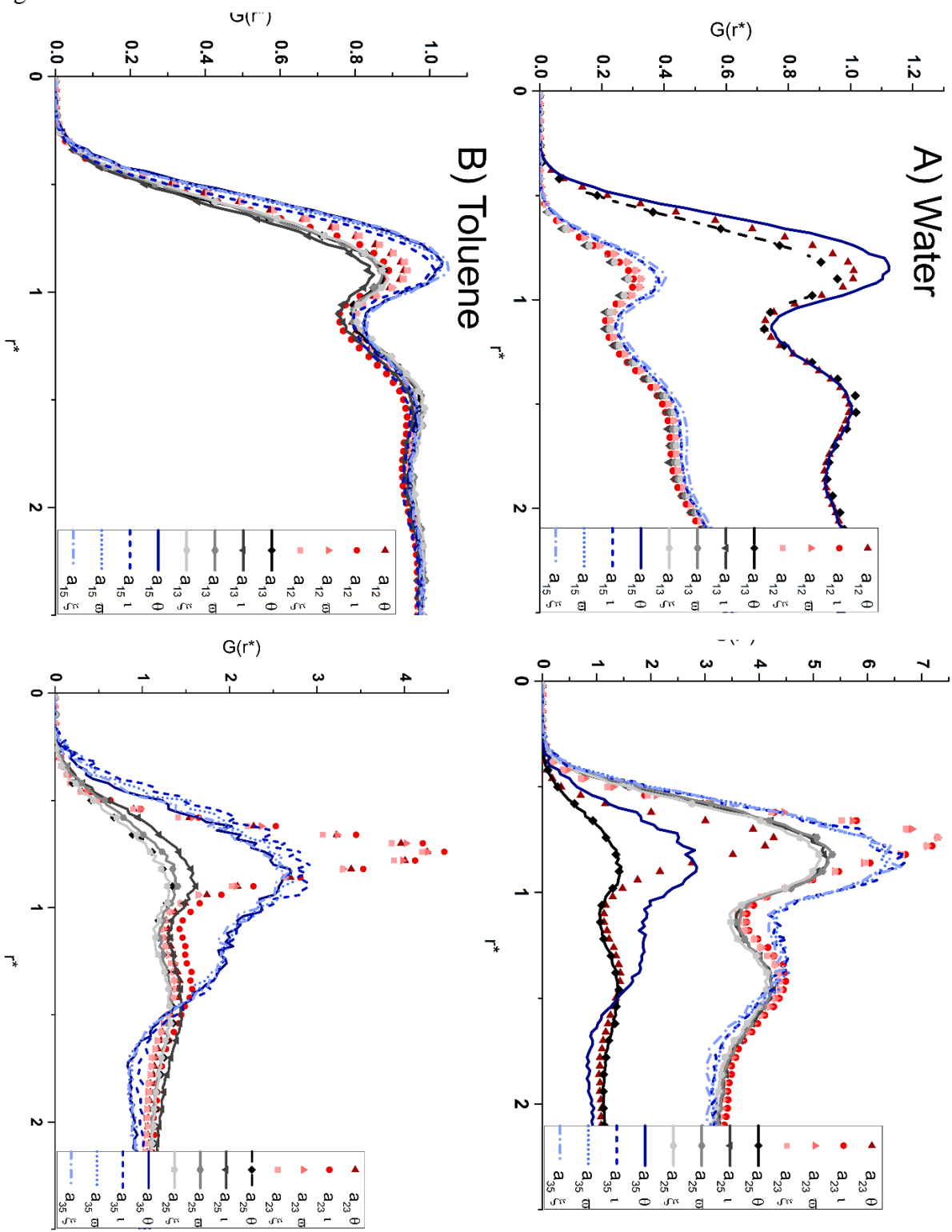

Figure 11

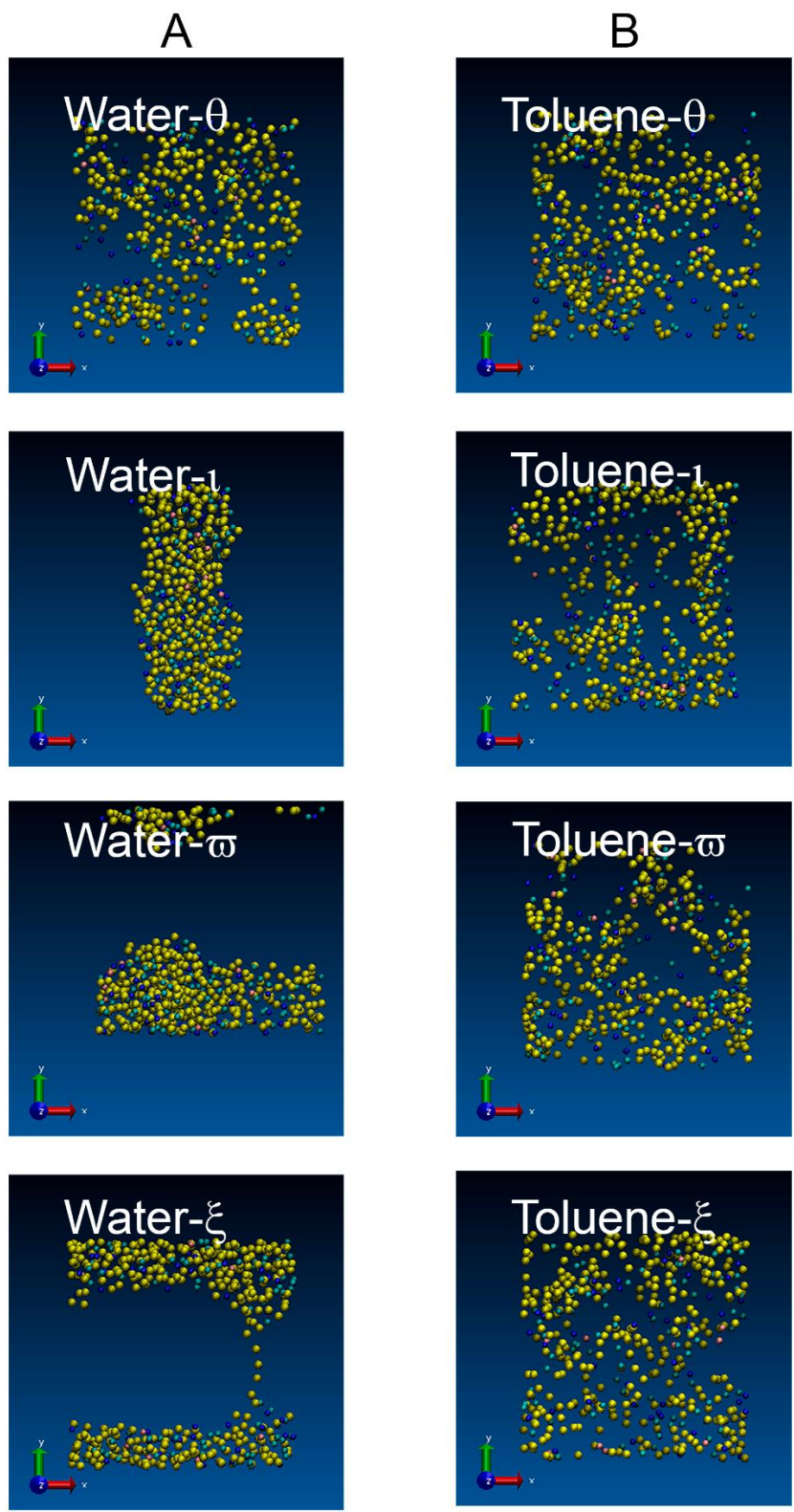